\documentclass[11pt,twoside]{book}
\usepackage{konkolyproc2}
\usepackage{longtable}
\usepackage{amsmath,amssymb}
\usepackage{graphicx}
\usepackage{lscape}
\usepackage{index}
\usepackage{natbib}
\usepackage{bigdelim}
\usepackage{multirow}
\makeindex

\begin{document}

\pagestyle{myheadings}
\setcounter{equation}{0}\setcounter{figure}{0}\setcounter{footnote}{0}
\setcounter{section}{0}\setcounter{table}{0}\setcounter{page}{1}
\markboth{Koll\'ath}{The unique dynamical system underlying RR Lyrae
pulsations}
\title{The unique dynamical system underlying RR Lyrae pulsations}
\author{Zolt\'an  Koll\'ath$^{1,2}$}
\affil{$^1$Konkoly Observatory, Research Centre for Astronomy and Earth Sciences, Hungarian Academy of Sciences, H-1121, Budapest, Konkoly Thege Mikl\'os \'ut 15-17, Hungary\\
$^2$University of West Hungary, Szombathely, Hungary}

\begin{abstract}  
Hydrodynamic models of RR Lyrae pulsation display a very rich behaviour. 
Contrary to earlier expectations, high order resonances play a crucial role 
in the nonlinear dynamics representing the interacting modes. 
Chaotic attractors can be found at different time scales: both in the 
pulsation itself and in the amplitude equations shaping the possible 
modulation of the oscillations. Although there is no one-to-one connection
between the nonlinear features found in the numerical models and the 
observed behaviour, the richness of the found phenomena suggests that the 
interaction of modes should be taken seriously in the study of the still 
unsolved puzzle of Blazhko effect. One of the main lessons of this complex 
system is that we should rethink the simple interpretation of the observed 
effect of resonances.
\end{abstract}

\vspace*{-7mm}
\section{Introduction}
\vspace*{-2mm}
Recent space- and ground-based observations of RR Lyrae stars have resulted 
in a golden age of stellar pulsation studies. A new level of complexity is 
uncovered; new pieces of the great puzzle of RR Lyrae variability have been 
found: e.g. unexpected pulsation modes and resonances. One of the most 
interesting features among the new findings is the period doubling visible 
in most of the Blazhko modulated stars \citep{szabo10}. Furthermore this 
effect is clearly the consequence of a high order (9:2) resonance of the 
fundamental mode with a strange 9th overtone \citep{kmsz11}. 

In addition, observations and hydrodynamic models suggest that a third 
mode may play an important role in the dynamics. Considering that the 
interaction of two modes leads to a rare and unique dynamical system, the 
interaction of the base system with additional modes may produce even more 
interesting behaviour.

The final question remains still unsolved: whether the very complex 
interaction of modes can solve the mystery of the Blazhko effect. 
Anyway, the dynamics of RR Lyrae pulsations provide useful lessons in 
general in relation to nonlinear processes in nature. In this paper we 
concentrate only one aspect of the complex dynamical system underlying 
RR~Lyrae pulsation: how the resonance appears in modulated solutions. 
This question is investigated both for the amplitude equations and 
hydrodynamic calculations.

\section{A rare chaotic attractor related to the 9:2 resonance}
\vspace*{-2mm}

\begin{figure}[!ht]
\includegraphics[width=1.0\textwidth]{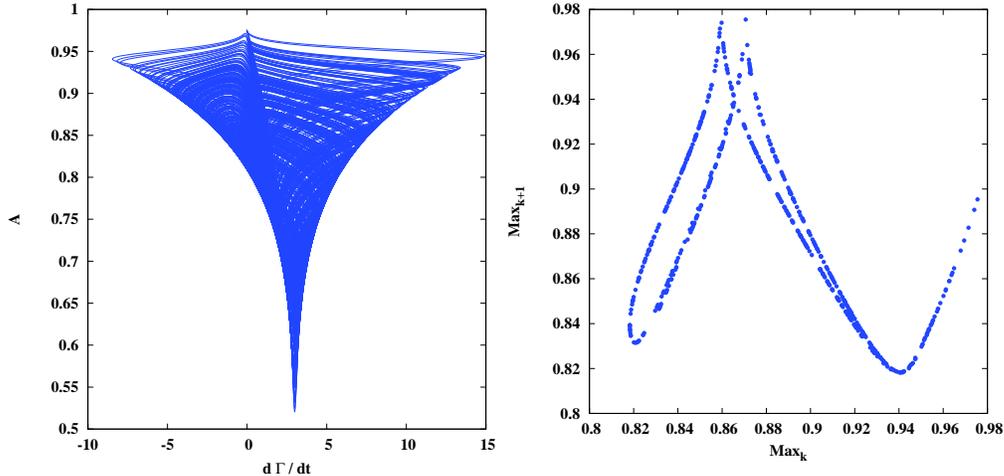}
\vspace*{-4mm}
\caption{The chaotic attractor of the resonant amplitude equations 
(left) and corresponding first return map (right).} 
\label{k-fig1} 
\end{figure}

The period doubling bifurcation found in the hydrodynamic models is a 
consequence of the resonant interaction of the 9th overtone with the 
fundamental mode. This effect can be understood by a simplified model 
of pulsations: if the solution of the amplitude equations with 
half-integer resonances is a nonzero fixed point (i.e. the amplitudes 
of both modes are constant) this system provides a natural description 
of period doubling. \citet{buchler-kollath} demonstrated, that the same 
amplitude equations have solutions that may explain the modulation of 
RR~Lyrae stars, furthermore this modulation can be chaotic. In this paper 
we use a simplified version of the resonant amplitude equations which 
is capable to present the major properties of the system:
\begin{eqnarray}
\frac{dA_0}{dt} &=& \big(1 - A_0^2 - r A_9^2\big) A_0 + c A_0^8 A_9^2 \cos(\Gamma) \nonumber \\
\noalign{\vskip 4pt}
\frac{dA_9}{dt} &=& \big(\kappa - A_9^2 - s A_0^2\big) A_9 + b A_0^9 A_9 \cos(\Gamma) \nonumber \\
\noalign{\vskip 4pt}
\frac{d\Gamma}{dt} &=& 2 \Delta - 9 c A_0^7 A_9^2 \sin(\Gamma) -2 b A_0^9 \sin(\Gamma),  \nonumber \\
&&
\label{eqs_ae}
\end{eqnarray}
where $A_0$, $A_9$ are the amplitudes of the fundamental mode and the 
9th overtone; $c$, $b$, $r$, $s$ and $\kappa$ are constant parameters. 
$\Gamma$ is the phase differences of the modes, its derivative gives the 
instantaneous frequency difference from the resonance: 
$d\Gamma / dt =2 \omega_0 - 9\omega_9$. The constant term which forces 
the system to deviate from the resonance is given by $\Delta$. 
Please note that we reduced the number of coefficients by setting the 
time scale to the growth rate of the fundamental mode and by normalizing 
the amplitudes.

The strange attractor of the system and the first return map calculated 
from the maxima of the fundamental mode amplitude variation are displayed 
in Fig.~\ref{k-fig1}. The parameters used for the figure are the 
following: $c=1.5$, $b=10$, $r=s=1.0$, $\kappa=-0.1$ and $\Delta=1.5$. 
This chaotic attractor represents a very rare type of dynamics: a double 
cusp map, where the two leaves of the attractor do not superpose 
exactly. The only known system of this type is the one proposed by 
\citet{rossler}. It is remarkable that the interaction of two 
modes results in such a rich dynamical phenomena.

We have to note that modulated solutions of the amplitude equations 
exist only for non-zero values of the offset parameter $\Delta$. 
When $\Delta$ is small, the resonance locks and a clean period doubled 
state is obtained. One of the most important pieces of information one 
can learn from Fig.~\ref{k-fig1} is that $d\Gamma / dt$ has a non zero 
mean and a wide distribution. It clearly demonstrates that in the 
Fourier spectra of such systems, broader and offset structures are 
expected at the half-integer frequencies instead of single peaks.

\vspace*{-2mm}
\section{Modulation and resonances in hydrodynamic models}
\vspace*{-2mm}

There exist period doubled solutions in RR Lyrae hydrodynamic models, 
but no model has been found with Blazhko-like modulation. However, 
\citet{sm12} found both phenomena in BL Herculis models, indicating 
that such behaviour is expected in hydrodynamic calculations. 
Moreover, it has been already demonstrated by \citet{pkm13} that period 
doubling coexists with chaotic and multimode pulsations in RR~Lyrae models.

\begin{figure}[!hb]
\includegraphics[width=1.0\textwidth]{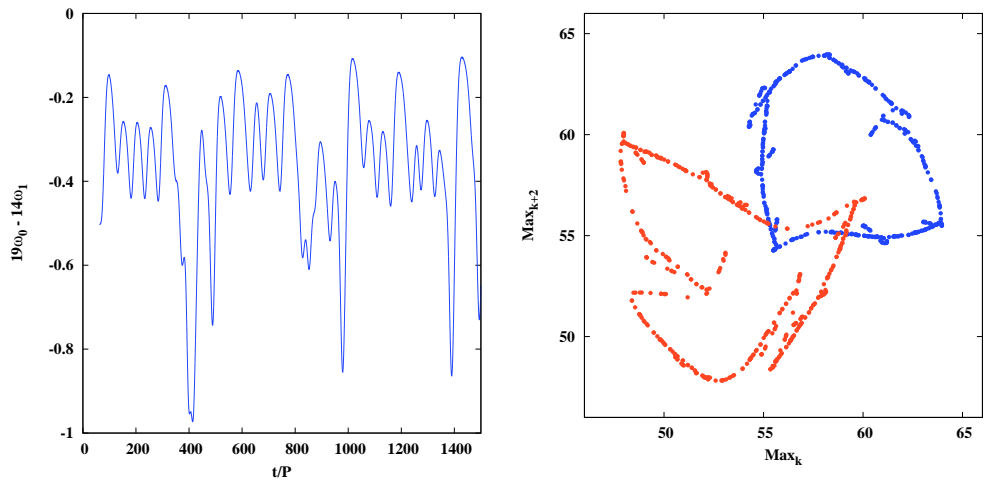}
\vspace*{-4mm}
\caption{The variation of the relative phase in a hydrodynamic model 
close to a 14:19 resonance (left) and the corresponding first return 
map of the radius maxima (right).} 
\label{k-fig2} 
\end{figure}

When a third mode appears in the interplay of oscillations, the number 
of possible solutions is increased. For example the interaction of 
the fundamental mode with the strange mode alters the stability properties 
of the fundamental mode and makes it more sensitive to other pulsation 
modes, e.g. our model calculations demonstrated that the period doubled 
version of the same oscillation now may lose its stability with respect 
to the first overtone. While in normal double mode scenario of the 
fundamental mode and the first overtone there is only one possible 
solution where both modes coexist, in the presence of the 9th overtone 
there is a possibility for additional solutions with a double mode 
character. Figure~2 in \citet{pkm13} displays the first return maps 
calculated from the radius variation of models representing this kind 
of behaviour.

The right panel in Figure \ref{k-fig2} in this paper displays a 
similar model. By plotting the maxima against the second previous 
maximum values, the period-doubling structure can be easily recognized 
(the two branches representing period doubling are plotted with 
different colours). The substructure of the individual branches are 
related to the effect of the first overtone which is in a close 14:19 
resonance with the fundamental mode. The agglomerations (7 in both 
branches) in the otherwise smooth return maps are the signs of the 
still active resonant interaction. In exact resonance the diagram reduces 
to 14 points in the plot. The instantaneous frequencies are calculated 
from the radius variation, and the offset from the resonance centre is 
plotted in the left panel of the figure. Similarly to the 9:2 resonant 
amplitude equations, an even higher order resonance can play a role in 
shaping the light variation. And more importantly, an offset from the 
resonance center results in a slight modulation of the fundamental mode 
amplitude and a nonzero mean in the $19\omega_0 - 14\omega_1$ off-resonance frequency difference.
   
\vspace*{-2mm}
\section{Conclusion}
\vspace*{-2mm}

It is demonstrated that high order resonances play an important role 
in the dynamics of RR Lyrae model pulsations. However, the exact, phase 
locked resonance is just one of the possible solutions. In general, 
when modulation emerges in a resonant system, the behaviour of the 
dynamics is significantly changed. The main characteristics of these 
systems are: 
\begin{itemize}
\item If the frequency offset from the resonance reaches a specific 
level, the phases are not locked to each other. Even then, fingerprints 
of the resonance are still observable in the resulting variations. 
\item The peaks in the Fourier transform of modulated  resonant systems 
do not coincide exactly with the resonant frequencies (e.g. the half 
integer ones), but there is an offset compared to the resonance centre. 
\item Broad structures dominate the spectrum instead of single resonant 
peaks. 
\end{itemize}

\vspace*{-5mm}
\section*{Acknowledgments}
\vspace*{-2mm}
This research has been supported by the NKFIH K-115709 grant and the
\linebreak European Community's FP7/2007-2013 Programme under grant
agreements no.~269194 (IRSES/ASK) and no.~312844 (SPACEINN).

\end{document}